\documentclass[12pt]{iopart}

\usepackage{graphicx}
\usepackage{dcolumn}
\usepackage{bm}

\usepackage[utf8]{inputenc}
\usepackage{caption}
\usepackage[T1]{fontenc}
\usepackage{etoolbox}
\usepackage[
backend=biber,
style=numeric,
sorting=none,
giveninits=true
]{biblatex}
\begin{document}

\title{Analysis of Thermal Grooving Effects on Vortex Penetration in Vapor-Diffused Nb$_3$Sn}

\author{Eric M. Lechner$^1$, Olga Trofimova$^1$, Jonathan W. Angle$^2$, Madison C. DiGuilio$^{3}$, Uttar Pudasaini$^1$}

\address{$^1$Thomas Jefferson National Accelerator Facility, Newport News, Virginia 23606}
\address{$^2$Pacific Northwest National Laboratory, Richland, Washington 99354}
\address{$^3$Tidewater Community College, Norfolk, Virginia 23510}
\ead{lechner@jlab.org, uttar@jlab.org}
\vspace{10pt}
\begin{indented}
\item[]November 27, 2024
\end{indented}

\begin{abstract}
While Nb$_3$Sn theoretically offers better superconducting radio-frequency cavity performance (\(Q_0\) and \(E_{acc}\)) to Nb at any given temperature, peak RF magnetic fields consistently fall short of the $\sim$400 mT prediction. The relatively rough topography of vapor-diffused Nb$_3$Sn is widely conjectured to be one of the factors that limit the attainable performance of Nb$_3$Sn-coated Nb cavities prepared via Sn vapor diffusion.  Here we investigate the effect of coating duration on the topography of vapor-diffused Nb$_3$Sn on Nb and calculate the associated magnetic field enhancement and superheating field suppression factors using atomic force microscopy topographies. It is shown that the thermally grooved grain boundaries are major defects which may contribute to a substantial decrease in the achievable accelerating field. The severity of these grooves increases with total coating duration due to the deepening of thermal grooves during the coating process.
\end{abstract}

%
%
%
%
%

Radio-frequency (RF) cavities are electromagnetic resonators capable of storing electromagnetic energy to accelerate charged particle beams in modern accelerators. Compared with normal conducting RF cavities, superconducting RF (SRF) cavities possess much smaller surface resistance, $R_s$, reducing the RF power dissipation in the cavity walls by 5-6 orders of magnitude compared to normal conducting materials. This low surface resistance allows SRF cavities to operate at high fields in continuous wave mode. Bulk niobium ($T_{c} \approx$ 9.2 K, $B_{c} \approx$ 200 mT, $B_{sh}$ $\approx$ 240 mT, and $\Delta$ $\approx$ 1.5 meV \cite{Desorbo1963EffectsOfONinNb,Ciovati2013WhereNextWithSRF,Matricon1967SuperheatingFields,Lechner2020ElectronTunneling}) is currently the material of choice to fabricate SRF cavities due to its superior superconducting properties among the pure elements and suitable mechanical properties to form complex structures. Over the past five decades, continual research and development has advanced Nb SRF cavity technology, determining performance limiting mechanisms, and developing adequate processes to mitigate them. 

Niobium cavities often require operating at $\sim 2$ K for optimal performance, which demands complicated cryogenic facilities, and is a major cost driver for SRF-based accelerators. Nb$_3$Sn is an alternative material with potential to surpass the state-of-art Nb cavity performance. Nb$_3$Sn possesses superior superconducting properties: $T_c \approx$ 18.3 K, $B_{c}$ $\approx$ 470 mT, $B_{sh}$ $\approx$ 400 mT, and $\Delta$ $\approx 3.4$ meV \cite{Godeke2006Nb3SnVariation,Posen2014AdvanceInNb3Sn,Keckert2019CriticalFieldsofNb3SnCavities,Moore1979EnergyGapsOfNb3SnV3SiNb3Ge}, all almost twice that of Nb. Theoretically, it promises higher accelerating gradient, quality factor, and operating temperature than bulk Nb. Nb$_3$Sn SRF cavities at 4.3 K can deliver similar performance to Nb cavities at 2 K. This would enable enormous cost savings for future SRF accelerators by simplifying cryogenic facilities and reducing their operating cost. Successful deployment of Nb$_3$Sn technology would usher in an era of high efficiency, conduction-cooled accelerators for basic science, medicinal and industrial applications \cite{Kephart2015CompactAccel4IndustrySociety,Barletta2010ReportOnCompactLightSources}.

Nb$_3$Sn cavities routinely operate at peak magnetic fields between $B_{c1}$ and $B_{sh}$, indicating $B_{c1}$ is not a limiting parameter for the operation in the metastable Meissner state. It has been shown using both DC and short high power pulses that a major limitation may be related to flux penetration at defects in the RF layer \cite{Posen2015RFDCFieldLimitsOfNbAndNb3Sn}. While Nb$_3$Sn theoretically offers superior peak accelerating fields, they are typically limited below 20 MV/m ($85$ mT)\cite{PosenHall2017Nb3SnSRFReview}. To date, the highest accelerating gradient at 4.2 K in vapor-diffused Nb$_3$Sn 1.3 GHz single cell cavities is 24 MV/m ($100$ mT)\cite{PosenAdvancesInNb3Sn24MVm2021}. An interesting observation from witness sample studies of that cavity showed unusually small surface roughness. 

The smoothness of the RF surface is one of the important aspects to achieve high-field cavity performance. Smoother is better. As such, an ideal surface is perfectly flat. Surface roughness on a superconductor exposed to an applied magnetic field introduces two vulnerabilities to the preservation of the Meissner state: magnetic field enhancement (MFE) and superheating field suppression (SFS). Magnetic field enhancement arises from the supercurrent screening of applied magnetic fields by rough surfaces where the magnetic field is locally increased. The most severely affected regions are marked by sharp edge/small radius of curvature topographic defects \cite{knobloch1999high,xu2016simulation,gurevich2006multiscale,kubo2015magnetic,shemelin2008magnetic}. The local field due to MFE can exceed the critical magnetic field of the material and cause regions of the defect to transit into the normal conducting state. These normal conducting regions expand and lead to thermal instability \cite{knobloch1999high,xu2016simulation,Xie2011QuenchSimulation,Xie2013QuenchAndFieldDependentSurfaceResistance}. Localized superheating field suppression, a weakened stability of the metastable Meissner state, arises due to nanoscale roughness \cite{aladyshkin2001best,kubo2015field,pack2020vortex}. This weakened stability results from the competition between the external magnetic field pushing a vortex into the superconductor and the image force from the surface expelling it. Topographic defects facilitate premature breakdown of the metastable Meissner state by enhancing the force from the external magnetic field pushing a vortex into the surface due to current crowding, a concept familiar to optimization of superconducting nanocircuits \cite{Clem2011CriticalCurrentNanocircuits,Hortensius2012CriticalCurrentReduction}, and reducing the force from the surface expelling it \cite{kubo2015field}. At \(B_{sh}\) the surface is absolutely unstable to nucleation of highly dissipative vortices that may contribute to thermal instability \cite{gurevich2008dynamics,gurevich2012superconducting,Carlson2021Analysis}. Recent analysis of N-doped and low temperature baked Nb showed that the preservation of B$_{sh}$ may play a role in experimentally attainable peak field \cite{Lechner2023TopographicEvolution,LechnerOxideDissoScenarios2024}.

\begin{figure}[!ht]
\includegraphics[width = 8.5 cm]{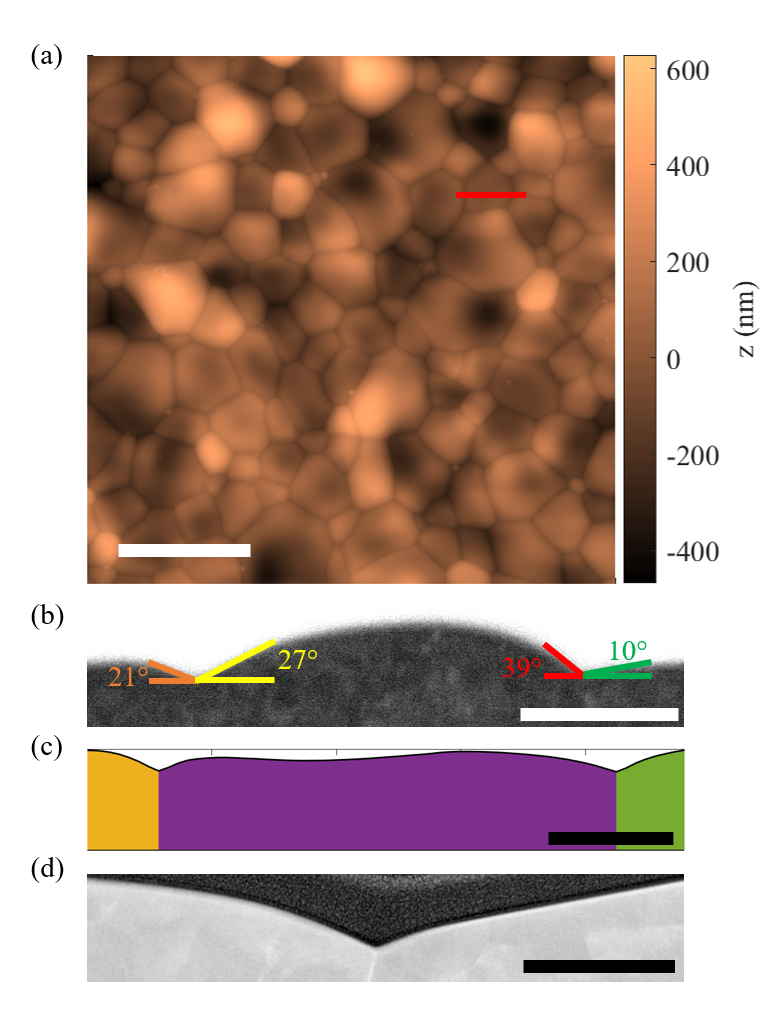}
\centering
\caption{(a) Representative TMAFM topography of a Nb$_3$Sn surface after 6 hours of coating time at $T = $ 1200 °C. The scale bar is 5 $\mu$m. (b) SE-SEM image of a cross-section through a grain. The scale bar is 1 $\mu$m. (c) Topographic line profile through the area marked by the red line in (a). Grains are marked by different colors beneath the topographic line profile. The scale bar is 500 nm. (d) HAADF-STEM image of a grain boundary groove cross-section. The scale bar is 200 nm.}
\captionsetup{justification=centering}
\label{Nb3SnRepresentativeTopograph}
\end{figure}

\begin{figure}[!ht]
\centering
\includegraphics[width = 8.5 cm]{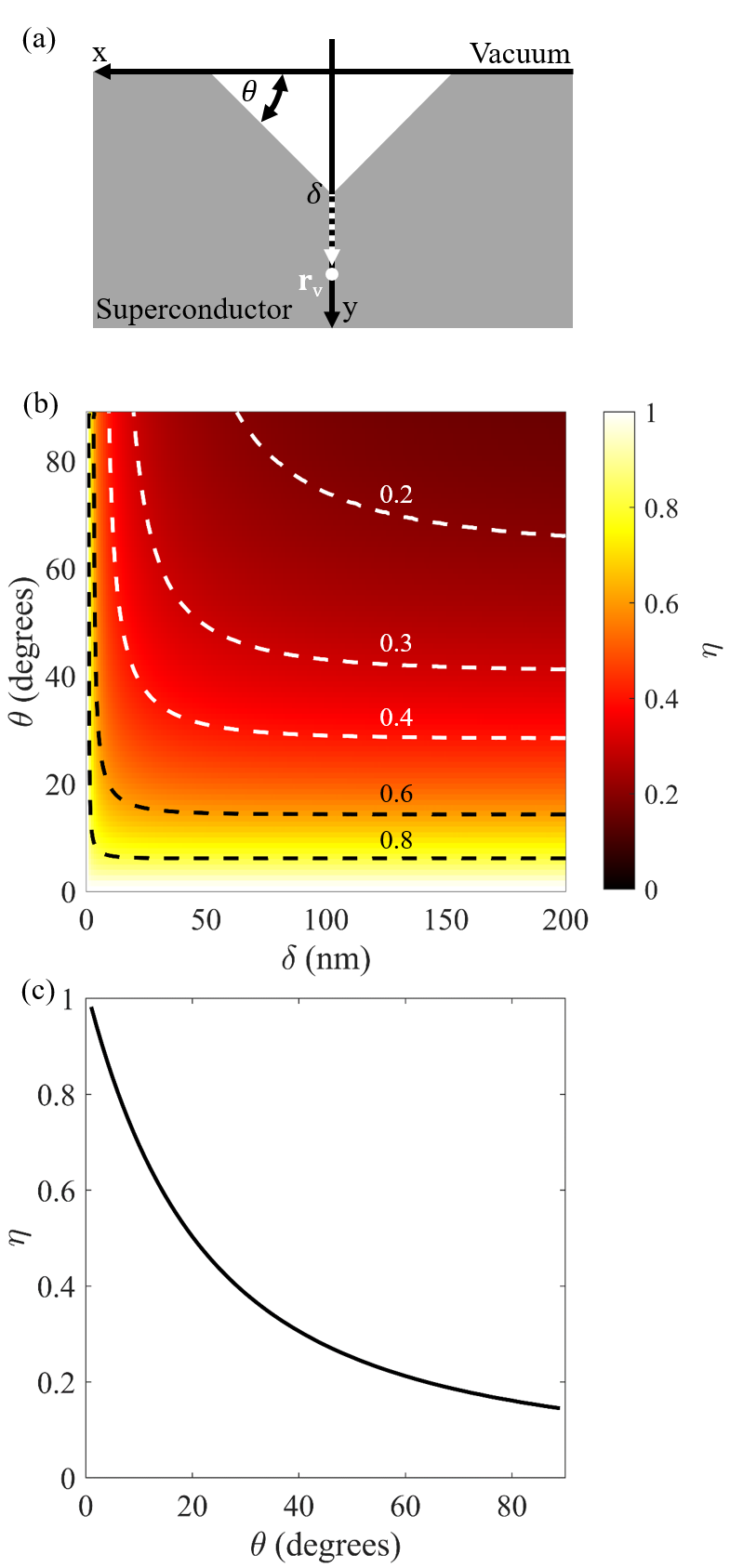}
\caption{(a) Triangular groove geometry used in the SFS model. (b) Color plot of $\eta(\delta,\theta)$ via Eq. \ref{eq5}. (c) $\eta(\theta)$ in the deep-groove limit. $\xi$ = 3 nm and $\lambda$ = 120 nm were used to represent Nb$_3$Sn.}
\captionsetup{justification=centering}
\label{FigSFSFLandscape}
\end{figure}

In this work the effect of coating duration on the topography of Sn vapor-diffused Nb$_3$Sn samples and its implications on SFS and MFE is investigated. Nb$_3$Sn coatings were prepared using a similar Sn vapor diffusion process used to coat Nb$_3$Sn on Nb cavities. The samples were prepared during a study to elucidate the growth mechanism of Nb$_3$Sn coating during the vapor diffusion process \cite{pudasaini2019growth}. This coating consists of two steps: nucleation and growth. Tin chloride is first evaporated at $\sim$500 °C, depositing a Sn film and
particles onto the Nb surface to mitigate non-uniformity in the coating by improving nucleation \cite{PudasainiInitialGrowthOfNb3Sn2019}. The Nb$_3$Sn growth temperature was 1200 °C, well above the temperature of 930 °C needed to exclusively form the Nb$_3$Sn phase. Topographic measurements were acquired using a Dimension Icon Atomic Force Microscope, Bruker, at the Applied Research Center Core Labs of College of William and Mary. All measurements were made in tapping mode (TMAFM), using scanasyst-air probes (Bruker) \cite{ScanasystAirProbe} at 512 $\times$ 512 pixels.
As grown vapor-diffused Nb$_3$Sn surfaces are topographically imperfect. A representative Nb$_3$Sn surface is shown in Fig. \ref{Nb3SnRepresentativeTopograph} (a). Grain boundaries host high slope angle grooves as highlighted by the secondary electron scanning electron microscopy (SE-SEM) image, line profile through the topography in (a) and high angle annular dark field scanning transmission electron microscopy (HAADF-STEM) shown in Fig. \ref{Nb3SnRepresentativeTopograph} (b-d). The surface is characterized by an undulating intragranular topography separated by V-shaped, grooved grain boundaries. The grooved grain boundaries are qualitatively consistent with the shape of thermal grooving via surface diffusion \cite{Mullins1957ThermalGrooving,Akyildiz2017ThermalGrooving}. The groove root, which should retain a self-similar shape during growth\cite{Mullins1957ThermalGrooving}, is sharp as shown in Fig. \ref{Nb3SnRepresentativeTopograph} (d). These topographic defects give rise to local MFE and SFS. The MFE and SFS factors, $\beta(\mathbf{r})$ and $\eta(\mathbf{r})$, are defined by $B(\mathbf{r})=\beta(\mathbf{r}) B_0$ and $B^*_{sh}(\mathbf{r})=\eta(\mathbf{r}) B_{sh}$, respectively. Here $B_0$ is the magnetic field far from the surface, $B(\mathbf{r})$ is the local magnetic field, $B_{sh}$ is the superheating field of a perfectly flat surface and $B^*_{sh}(\mathbf{r})$ is the geometrically modified superheating field. The MFE and SFS factors were calculated using the perfect electrical conductor model \cite{Lechner2023TopographicEvolution} and considering a triangular groove geometry similar to the one proposed by Kubo \cite{kubo2015field} to calculate SFS, as shown in Fig. \ref{FigSFSFLandscape} (a). In this simple model, we consider the competition between the force of the external magnetic field, $\mathbf{F_M}$, driving a vortex into the surface and the image vortex force, $\mathbf{F_S}$, expelling it from the surface. The force from the external magnetic field is given by

\begin{equation}
\mathbf{F_M}=\mathbf{J_M} \times \phi_0 \hat{\mathbf{z}}
\label{eq1}
\end{equation}
and the force from the surface is determined by the interaction with the image vortex given by 
\begin{equation}
\mathbf{F_S}=\mathbf{J_I} \times \phi_0 \hat{\mathbf{z}}.
\label{eq2}
\end{equation}
where $\mathbf{J_M}=\nabla\times \mathbf{B}/\mu_0$ is determined by solution of the London equation of an infinitely long triangular groove defect aligned in the z direction. In this case the solution, $\mathbf{B}(x,y)$, lies only in the z direction leading to the governing equation
\begin{equation}
\nabla^2 B_z = \frac{1}{\lambda^2}\ B_z.
\label{MagneticVectorPotentialWithin}
\end{equation}
where $\lambda$ is the London penetration depth with the boundary conditions $\mathbf{B}(x,y=h(x))=B_0\hat{z}$, $\mathbf{B}(x,y=\infty)=0$ and $\nabla B_z\cdot \mathbf{\hat{x}}=0$. Where $h(x)$ defines the surface contour. $\mathbf{J_I}=-\nabla\Phi_I$, where $\Phi_I$ is the current density scalar potential of the image vortex calculated via conformal mapping

\begin{equation}
\frac{2\pi\mu_0\lambda^2}{\phi_0}\Phi_{I}(x,y)=\tilde{\Phi}_{I}(x,y)=\textnormal{Re}\left(-i{\ln(w-w^*_0)}\right)|_{w=F^{-1}(x,y)}.
\label{eqJSPAV}
\end{equation}
$w$ is the complex coordinate, $w_0^*$ is the complex conjugate of the vortex nucleation position in the $w$-plane which preserves the boundary condition of zero normal current density to the surface, and $F^{-1}$ is the inverse map between $w$ and real coordinate calculated numerically and described in detail elsewhere\cite{kubo2015field}. Here the coherence length, $\xi$, is used as the cutoff length scale in the London theory \cite{Bean1964SurfaceBarrier,kubo2015field}.
When the sum of the forces on the vortex vanishes, $\mathbf{F_S} + \mathbf{F_M}=\mathbf{0}$, the Bean-Livingston barrier becomes unstable to vortex penetration \cite{kubo2015field} which leads to the following expression 

\begin{equation}
B_{sh}^*=\frac{2\xi|-\nabla\tilde{\Phi}_{I}(\mathbf{r_v})|}{\epsilon(\mathbf{r_v})}B_{sh}.
\label{eq4}
\end{equation}
where $B_{sh} = \phi_0/4\pi\lambda\xi$ is the superheating field of a perfectly flat surface in the London theory, $\epsilon(\mathbf{r})=|\mathbf{J(r)}|/J_0$ is the local current density enhancement factor, and $\mathbf{r_v}$ is the vortex nucleation position given by $\mathbf{r_v}=(\delta+\xi)\mathbf{\hat{y}}$ for this geometry. This defines the SFS factor 
\begin{equation}
\eta(\xi,\lambda,\delta,\theta)=\frac{2\xi|-\nabla\tilde{\Phi}_{I}(\mathbf{r_v})|}{\epsilon(\mathbf{r_v})}.
\label{eq5}
\end{equation}
which is computed numerically. This model predicts a superheating field suppression factor based on the slope angle of the groove, $\theta$, the depth of the groove, $\delta$, the superconductor’s coherence length, $\xi$, and the London penetration depth, $\lambda$. This model spans from the nanoscale limit considered by Kubo \cite{kubo2015field} to the deep groove geometry considered by Buzdin and Daumens \cite{Buzdin1998EMPinningOfVorticesOnDefects} and can be used between the two limits where many of the defects in the Nb$_3$Sn system find themselves. Using Eq. \ref{eq5} and estimating $\xi \approx$  3 nm and $\lambda \approx$ 120 nm for Nb$_3$Sn\cite{posen2015understanding,Liarte2017TheoreticalEstimatesOfMaximumFields,Hall2017PhDThesis}, $\eta(\delta,\theta)$ is plotted in Fig. \ref{FigSFSFLandscape} (b). The superconducting coherence length and penetration depth values used are qualitatively useful for evaluating the effect of surface roughness on the superheating field for other materials\cite{valente2016superconducting}. The surface in Fig. \ref{FigSFSFLandscape} (b) shows that the SFS factor can be quickly degraded by defects of only a few 10's of nanometers deep and slope angles exceeding 10°. In the limit of deep grooving, the SFS factor ceases to change with increasing depth of the groove, which is consistent with the deep groove limit \cite{Buzdin1998EMPinningOfVorticesOnDefects}. The slope angle dependent deep-groove limit SFS factors for this model are plotted in Fig. \ref{FigSFSFLandscape} (c).

\begin{figure*}
\centering
\includegraphics[width=1.0\textwidth]{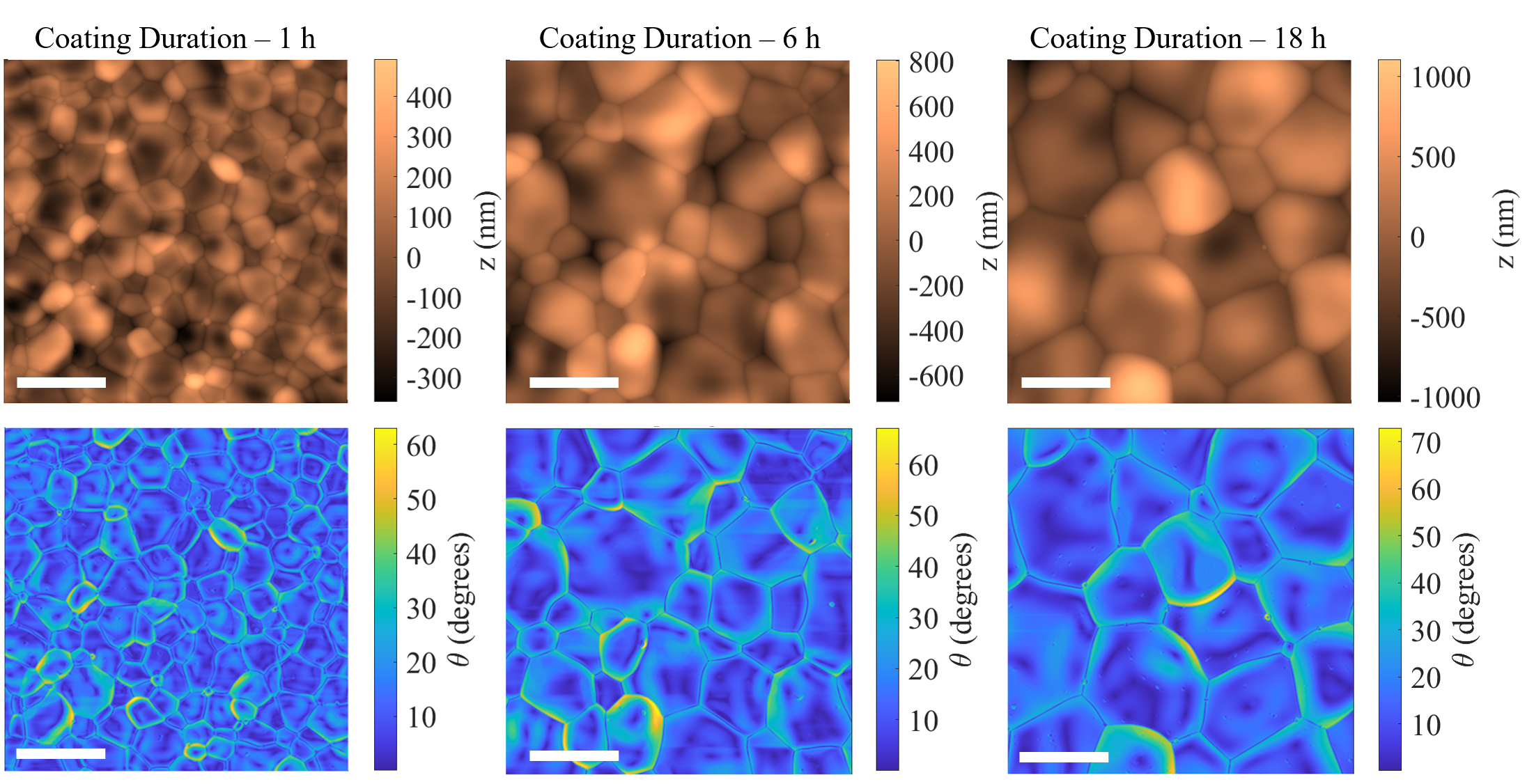}
\caption{Representative TMAFM topographies (upper panels) and slope angles (lower panels) showing the evolution of Nb$_3$Sn topography with coating duration. The scale bars are 5 $\mu$m.}
\captionsetup{justification=centering}
\label{RepresentativeCoatingTimeTopography}
\end{figure*}

\begin{figure*}
\centering
\includegraphics[width=1.0\textwidth]{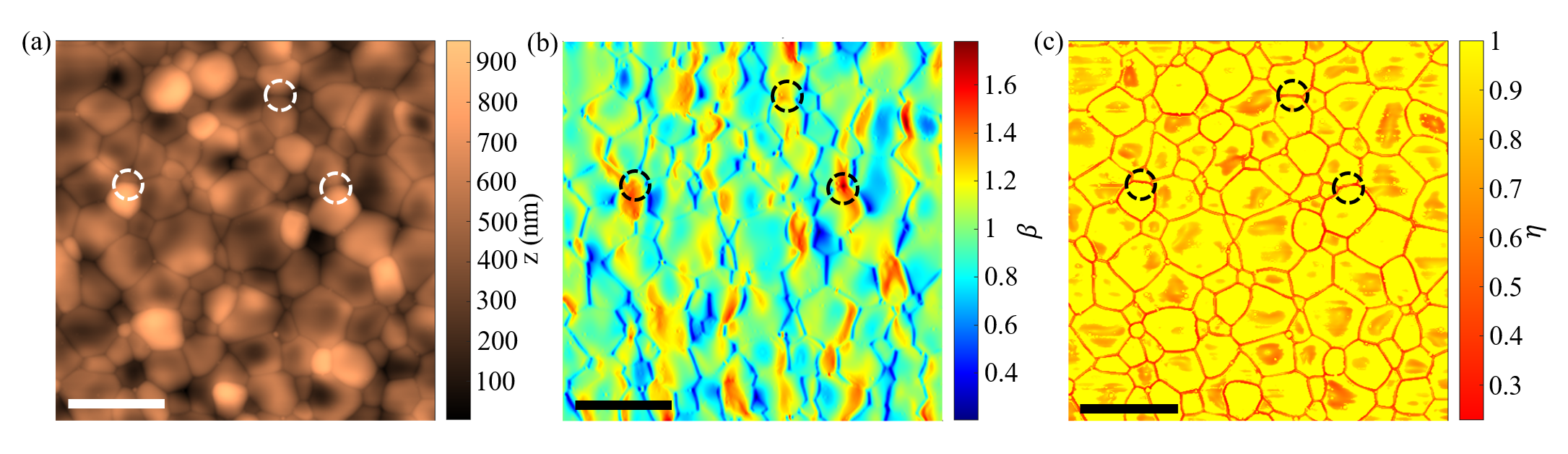}
\caption{(a) Representative topography of a Nb$_3$Sn sample coated at 1200 °C for 3 hours. (b) MFE factor map of the topography used in (a). (c) SFS factor map of the topography used in (a). Dashed circles indicate some of the regions on the surface where SFS and MFE coincide. The scale bars are 5 $\mu$m.}
\captionsetup{justification=centering}
\label{FigRepresentativeMFESFS}
\centering
\end{figure*}

The topographic evolution of Nb$_3$Sn with coating durations between 1 to 78 hours was examined. Ten topographic measurements sampled randomly across the surface of each sample were made at 20 $\mu$m $\times$ 20 $\mu$m for samples coated up to 18 h and at 25 $\mu$m $\times$ 25 $\mu$m for 60 h and 78 h. Representative TMAFM images are shown in Fig. \ref{RepresentativeCoatingTimeTopography}. Extended coatings develop on average larger grains and increased peak-to-valley distance as indicated by the z color scale in Fig. \ref{RepresentativeCoatingTimeTopography}. Local slope angles can be very large, exceeding 60° in some cases. As shown in Fig. \ref{RepresentativeCoatingTimeTopography}, the largest slope angles form exclusively between particularly tall grains and much shorter ones. As measured by TMAFM, large slope angles tend to form a slope-angled step geometry but could have more complex geometry not apparent to AFM \cite{Lee2019AtomicScaleAnalysesofNb3Sn}. Magnetic field enhancement factor maps are calculated using the methods outlined in previous work \cite{Lechner2023TopographicEvolution}. The SFS factor maps are calculated similar to that in \cite{Lechner2023TopographicEvolution} by determining $\delta(\mathbf{r})$ and $\theta(\mathbf{r})$ as inputs to Eq. \ref{eq5}. We estimate $\delta(\mathbf{r})$ by taking the difference between the topography and a plane formed by a first order, 11-pixel frame, Savitzky-Golay type\cite{kuo1991multidimensional,savitzky1964smoothing} smoothing of the topography. This serviceable estimate of $\delta(\mathbf{r})$ tends to underestimate $\delta$ due to the nature of the smoothing procedure on the complex topography of Nb$_3$Sn. MFE and SFS factor maps derived from the topography in Fig. \ref{FigRepresentativeMFESFS} (a) are shown in Fig. \ref{FigRepresentativeMFESFS} (b) and (c). As shown in Fig. \ref{FigRepresentativeMFESFS} (c), the grain boundary grooves host the most severely degraded superheating field suppression factors. Considering the combined effect of SFS and MFE on maximum supportable field,  $B^*_{max}(\mathbf{r})=(\eta/\beta)B_{max}$, the SFS factors are at least $\sim2\times$ more severe than the MFE factors observed in Fig. \ref{FigRepresentativeMFESFS} (b) which may account for a substantial reduction in peak supportable magnetic field in Nb$_3$Sn films since only a small fraction of vortex-active grain boundaries are required to increase surface resistance substantially\cite{Carlson2021Analysis}.

\begin{figure*}
\centering
\includegraphics[width=1.0\textwidth]{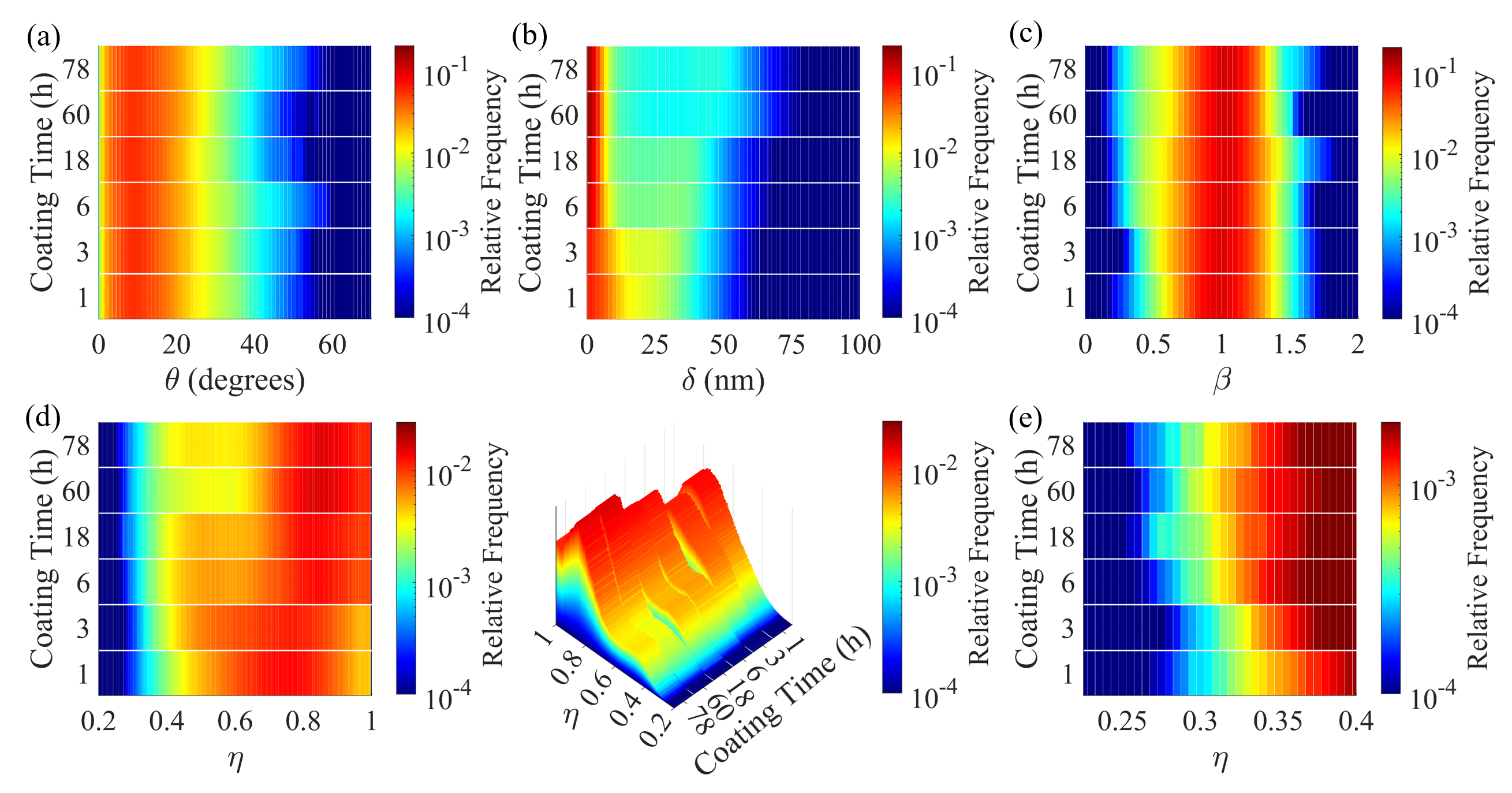}
\caption{(a) Relative frequencies of local slope angle, $\theta$, (b) deviation between topography and background plane, $\delta$, (c) local MFE factors, $\beta$, (d) local SFS factors, $\eta$, and a tilted view showcasing the low-$\eta$ bifurcation, and (e) a zoom-in of the low-$\eta$ SFS factors highlighting the effect of deepening thermal grooves with coating duration. The relative frequency cutoff of $10^{-4}$ corresponds to the significance of a 630 $\times$ 630 nm$^2$ patch for the 20 $\mu$m $\times$ 20 $\mu$m data sets.}
\captionsetup{justification=centering}
\label{SHFSFMFECoatingTime} 
\end{figure*}

The relative frequencies of parameters extracted from the AFM topographies are shown in Fig. \ref{SHFSFMFECoatingTime}. The distribution of $\theta(\mathbf{r})$ and $\delta(\mathbf{r})$ is shown in Fig. \ref{SHFSFMFECoatingTime} (a) and (b). No discernible trend with coating duration is observed in the distribution of $\theta(\mathbf{r})$, while the depth of the grooves, represented by the high-$\delta$ branch, tends to increase with increasing coating duration, as shown in Fig. \ref{SHFSFMFECoatingTime} (b). These observations are qualitatively consistent with the thermal grooving of grain boundaries where the dihedral angle is often assumed not to change \cite{Mullins1957ThermalGrooving} or expected to change very little \cite{Zhang2002ThermalGroovingChangingAngle} during high temperature annealing. The distribution of peak local MFE factors, shown in Fig. \ref{SHFSFMFECoatingTime} (c), also does not show any discernible trend as coating duration increases, which is likely due to the self-similar growth of thermal grooves. Peak MFE factors fall between 1.5 and 2 in agreement with previous estimates of MFE factors on Nb$_3$Sn \cite{Porter2016SurfaceRoughnessNb3Sn}. It is worth reiterating that the MFE factors obtained here were calculated in the framework of the perfect electrical conductor model, which assumes the limit $\lambda \rightarrow 0$, however the length scale of the defects is in fact comparable to the London penetration depth of Nb$_3$Sn of $\sim$ 120 nm. The comparative length scale effectively reduces the severity of MFE near sharp edges which would otherwise diverge in the perfect electrical conductor model. The SFS factors distributions shown in Fig. \ref{SHFSFMFECoatingTime} (d) presents a merged bimodal distribution that bifurcates as coating duration increases. This bifurcation is also reflected in the $\delta$ distribution in Fig. \ref{SHFSFMFECoatingTime} (b). The low-$\eta$ branch, related to the grain boundary grooving process, widens and extends toward lower-$\eta$ values and lower relative frequency. The extension to lower-$\eta$ values is due to the increasing depth of grain boundary grooves, while its decrease in relative frequency is due to the reduced sampling of grain boundaries in the scan areas caused by grain growth during extended coating duration. A zoom-in on low-$\eta$ values is presented in Fig. \ref{SHFSFMFECoatingTime} (e), which shows that the low-$\eta$ distribution reaches from $\eta\sim$0.3 for short coating times to $\eta\sim$0.25 for longer coating times. Despite the deepening of grooves in Fig. \ref{SHFSFMFECoatingTime} (b), the SFS factors stabilize and do not further decrease. This is due to the invariance of superheating field suppression factors on groove depth in the deep-groove limit as shown in Fig. \ref{FigSFSFLandscape} (b). Considering a theoretical superheating field of $\approx 400$ mT  for Nb$_3$Sn and the value of superheating field suppression factors alone, this would correspond to accelerating fields from $\sim$25-30 MV/m (107-128 mT) in TESLA-shaped cavities \cite{Aune2000TESLACavities}. Additional loss in peak field may be observed if MFE and SFS coincide. Such locations, some of which are circled in Fig. \ref{FigRepresentativeMFESFS} (a-c), may further reduce the peak accelerating field by a factor of $\sim$ 1.7, yielding $\sim$ 12-18 MV/m (51-77 mT). While this is in qualitative agreement with the field limitation observed in Nb$_3$Sn cavities, we expect that the London model estimation of the SFS factors represents a lower bound of $\eta$ due to non-locality of the BCS current-field relation. Additional causes of performance degradation are likely concurrent. Performance limiting mechanisms may take the form of Sn droplets \cite{JIANG2024optimizationNb3SnSRF}, thin patchy regions \cite{Pudasaini2018SurfaceStudiesNb3SnCoatedSamples}, reduced thermal conductivity \cite{CODY1964ThermalConductivityOfNb3Sn} or weak links caused by non-stoichiometry at grain boundaries\cite{Sandim2013Nb3SnGrainBoundarySegregation,LEE2020Nb3SnGrainBoundarySegregation,Carlson2021Analysis}. Recent calculations suggest that thermal conductivity is not currently a limiting factor in typically grown, few-$\mu$m thick films \cite{kulyavtsev2021simulations}. Based on this, these topographic defects may account for a substantial degradation of peak-field in dense and stoichiometric Nb$_3$Sn thin films grown via Sn vapor diffusion. If possible, groove slope angles and groove depths should be minimized to preserve the superheating field as shown in Fig. \ref{FigSFSFLandscape} (b). Coating duration should be optimized to minimize further degradation of SFS and MFE factors via thermal grooving. However, the coating must satisfy other conditions, like a minimum film thickness and uniformity which may not necessarily coincide with shortened coating duration.

In summary, we have investigated the effects of coating growth duration on the topography of vapor-diffused Nb$_3$Sn grown at 1200 °C. Within the perfect electrical conductor model and superheating field suppression model based on the London theory, we have shown that thermal grooves are particularly severe for SFS, and the effects of MFE are non-negligible but comparatively smaller. We expect improved estimations of SFS to be obtained from the Ginzburg-Landau theory using realistic superconducting properties, but this method quickly becomes computationally expensive for high $\lambda/\xi$ material \cite{pack2020vortex}. These thermal grooves should act as gates for vortices to enter the superconductor and limit achievable fields. It is imperative to seek processes that reduce groove slope angles and groove depths to mitigate performance degradation due to geometrical superheating field suppression. It may be possible to modify or inhibit the thermal grooving process via surface energy anisotropy \cite{rabkin2006MograinboundaryGroovingInMoBicrystals} or introducing impurities \cite{IWASAKI1997MolecularDynamicsGBGrooving}, however, care must be taken not to reduce the coherence length substantially, making the film even more susceptible to superheating field suppression \cite{kubo2015field,Buzdin1998EMPinningOfVorticesOnDefects}. In contrast to our present investigation, previous work showed that Nb$_3$Sn grown at different temperatures presented different distributions of MFE \cite{Pudasaini2023SRF23SurfaceRoughnessNb3Sn}. This suggests that growth temperature may have a sizable effect on the surface morphology. Exploring the effect of growth temperature as well as comparing RF performance with topography of witness samples is a natural next step. Additional post processing may be possible. Electropolishing \cite{Pudasaini2018ElectrochemicalFinishing}, oxypolishing \cite{Pudasaini2017PostProcessingNb3Nn,Porter2017EffectivenessChemicalTreatment}, and centrifugal barrel polishing \cite{Viklund2024Nb3SnCBP} may offer avenues to reduce the effects of surface roughness.

\section{Acknowledgments}
This material is based upon work supported by the U.S. Department of Energy, Office of Science, Office of Nuclear Physics under contract DE-AC05-06OR23177. This work was supported in part by the U.S. Department of Energy, Office of Science, Office of Workforce Development for Teachers and Scientists (WDTS) under the Science Undergraduate Laboratory Internships (SULI) program as well as the U.S. National Science Foundation Research Experience for Undergraduates
at Old Dominion University Grant No. 2348822. We gratefully acknowledge G. Eremeev and M. Kelley for their support to fabricate samples used in this work. We are grateful to C. Reece for insightful discussions.
\section{Data Availability}
The data that support the findings of this study are available from the corresponding author upon reasonable request.
\printbibliography
\end{document}